\documentclass[12pt]{article}
 
\def\datetitle{October 27, 1999}
\def\datehead{October 27, 1999}
 
\makeatletter
 
\@addtoreset{equation}{section}

\begin{document}

    \def\@evenhead{\rm\small\thepage\hfil\datehead}
    \def\@oddhead{\datehead\hfil\rm\small\thepage}

%%%%%%%%%%%%%%%%%%%%%%%%%%%%%%%%%%%%%%%%%%%%%%%%%%%%%%%%%%%%%%%%%%%%%%%%%%%%%
%
%  Redeclaration of \makeatletter; no @-expressions may be used from now on
%
%%%%%%%%%%%%%%%%%%%%%%%%%%%%%%%%%%%%%%%%%%%%%%%%%%%%%%%%%%%%%%%%%%%%%%%%%%%%%
 
\makeatother
 
%
% Symbols (Springer)
%
 
\def\bbbr{{\rm I\!R}} %reelle Zahlen
\def\bbbn{{\rm I\!N}} %natuerliche Zahlen
\def\bbbm{{\rm I\!M}}
\def\bbbh{{\rm I\!H}}
\def\bbbf{{\rm I\!F}}
\def\bbbk{{\rm I\!K}}
\def\bbbp{{\rm I\!P}}
\def\bbbone{{\mathchoice {\rm 1\mskip-4mu l} {\rm 1\mskip-4mu l}
{\rm 1\mskip-4.5mu l} {\rm 1\mskip-5mu l}}}
\def\bbbc{{\mathchoice {\setbox0=\hbox{$\displaystyle\rm C$}\hbox{\hbox
to0pt{\kern0.4\wd0\vrule height0.9\ht0\hss}\box0}}
{\setbox0=\hbox{$\textstyle\rm C$}\hbox{\hbox
to0pt{\kern0.4\wd0\vrule height0.9\ht0\hss}\box0}}
{\setbox0=\hbox{$\scriptstyle\rm C$}\hbox{\hbox
to0pt{\kern0.4\wd0\vrule height0.9\ht0\hss}\box0}}
{\setbox0=\hbox{$\scriptscriptstyle\rm C$}\hbox{\hbox
to0pt{\kern0.4\wd0\vrule height0.9\ht0\hss}\box0}}}}
\def\bbbe{{\mathchoice {\setbox0=\hbox{\smalletextfont e}\hbox{\raise
0.1\ht0\hbox to0pt{\kern0.4\wd0\vrule width0.3pt
height0.7\ht0\hss}\box0}}
{\setbox0=\hbox{\smalletextfont e}\hbox{\raise
0.1\ht0\hbox to0pt{\kern0.4\wd0\vrule width0.3pt
height0.7\ht0\hss}\box0}}
{\setbox0=\hbox{\smallescriptfont e}\hbox{\raise
0.1\ht0\hbox to0pt{\kern0.5\wd0\vrule width0.2pt
height0.7\ht0\hss}\box0}}
{\setbox0=\hbox{\smallescriptscriptfont e}\hbox{\raise
0.1\ht0\hbox to0pt{\kern0.4\wd0\vrule width0.2pt
height0.7\ht0\hss}\box0}}}}
\def\bbbq{{\mathchoice {\setbox0=\hbox{$\displaystyle\rm Q$}\hbox{\raise
0.15\ht0\hbox to0pt{\kern0.4\wd0\vrule height0.8\ht0\hss}\box0}}
{\setbox0=\hbox{$\textstyle\rm Q$}\hbox{\raise
0.15\ht0\hbox to0pt{\kern0.4\wd0\vrule height0.8\ht0\hss}\box0}}
{\setbox0=\hbox{$\scriptstyle\rm Q$}\hbox{\raise
0.15\ht0\hbox to0pt{\kern0.4\wd0\vrule height0.7\ht0\hss}\box0}}
{\setbox0=\hbox{$\scriptscriptstyle\rm Q$}\hbox{\raise
0.15\ht0\hbox to0pt{\kern0.4\wd0\vrule height0.7\ht0\hss}\box0}}}}
\def\bbbt{{\mathchoice {\setbox0=\hbox{$\displaystyle\rm
T$}\hbox{\hbox to0pt{\kern0.3\wd0\vrule height0.9\ht0\hss}\box0}}
{\setbox0=\hbox{$\textstyle\rm T$}\hbox{\hbox
to0pt{\kern0.3\wd0\vrule height0.9\ht0\hss}\box0}}
{\setbox0=\hbox{$\scriptstyle\rm T$}\hbox{\hbox
to0pt{\kern0.3\wd0\vrule height0.9\ht0\hss}\box0}}
{\setbox0=\hbox{$\scriptscriptstyle\rm T$}\hbox{\hbox
to0pt{\kern0.3\wd0\vrule height0.9\ht0\hss}\box0}}}}
\def\bbbs{{\mathchoice
{\setbox0=\hbox{$\displaystyle     \rm S$}\hbox{\raise0.5\ht0\hbox
to0pt{\kern0.35\wd0\vrule height0.45\ht0\hss}\hbox
to0pt{\kern0.55\wd0\vrule height0.5\ht0\hss}\box0}}
{\setbox0=\hbox{$\textstyle        \rm S$}\hbox{\raise0.5\ht0\hbox
to0pt{\kern0.35\wd0\vrule height0.45\ht0\hss}\hbox
to0pt{\kern0.55\wd0\vrule height0.5\ht0\hss}\box0}}
{\setbox0=\hbox{$\scriptstyle      \rm S$}\hbox{\raise0.5\ht0\hbox
to0pt{\kern0.35\wd0\vrule height0.45\ht0\hss}\raise0.05\ht0\hbox
to0pt{\kern0.5\wd0\vrule height0.45\ht0\hss}\box0}}
{\setbox0=\hbox{$\scriptscriptstyle\rm S$}\hbox{\raise0.5\ht0\hbox
to0pt{\kern0.4\wd0\vrule height0.45\ht0\hss}\raise0.05\ht0\hbox
to0pt{\kern0.55\wd0\vrule height0.45\ht0\hss}\box0}}}}
 
%
% note: changed \sans to \sf for LaTeX
%
 
\def\bbbz{{\mathchoice {\hbox{$\sf\textstyle Z\kern-0.4em Z$}}
{\hbox{$\sf\textstyle Z\kern-0.4em Z$}}
{\hbox{$\sf\scriptstyle Z\kern-0.3em Z$}}
{\hbox{$\sf\scriptscriptstyle Z\kern-0.2em Z$}}}}
 
%
%  Theorems and such
%
\newtheorem{theorem}{Theorem}[section]          % Numbering by sections
\newtheorem{lemma}[theorem]{Lemma}              % Number all in one sequence
\newtheorem{proposition}[theorem]{Proposition}
\newtheorem{corollary}[theorem]{Corollary}
\newtheorem{definition}[theorem]{Definition}
\newtheorem{conjecture}[theorem]{Conjecture}
\newtheorem{claim}[theorem]{Claim}
\newtheorem{observation}[theorem]{Observation}
\def\proof{\par\noindent{\it Proof.\ }}
\def\reff#1{(\ref{#1})}
%
% Springer symbols for numeral sets
%
\let\zed=\bbbz % \def\zed{{\hbox{\specialroman Z}}}
\let\szed=\bbbz % \def\szed{{\hbox{\sevenspecialroman Z}}}
\let\IR=\bbbr % \def\IR{{\hbox{\specialroman R}}}
\let\R=\bbbr % \def\IR{{\hbox{\specialroman R}}}
\let\sIR=\bbbr % \def\sIR{{\hbox{\sevenspecialroman R}}}
\let\IN=\bbbn % \def\IN{{\hbox{\specialroman N}}}
\let\IC=\bbbc % \def\IC{{\hbox{\specialroman C}}}

\def\nl{\medskip\par\noindent}
 
\def\scrb{{\cal B}}
\def\scrg{{\cal G}}
\def\scrf{{\cal F}}
\def\scrl{{\cal L}}
\def\scrr{{\cal R}}
\def\scrt{{\cal T}}
\def\pfin{{\cal S}}
\def\prob{M_{+1}}
\def\cql{C_{\rm ql}}
\def\bydef{:=}
\def\qed{\hbox{\hskip 1cm\vrule width6pt height7pt depth1pt \hskip1pt}\bigskip}
\def\remark{\medskip\par\noindent{\bf Remark:}}
\def\remarks{\medskip\par\noindent{\bf Remarks:}}
\def\example{\medskip\par\noindent{\bf Example:}}
\def\examples{\medskip\par\noindent{\bf Examples:}}
\def\nonexamples{\medskip\par\noindent{\bf Non-examples:}}

\newenvironment{scarray}{
          \textfont0=\scriptfont0
          \scriptfont0=\scriptscriptfont0
          \textfont1=\scriptfont1
          \scriptfont1=\scriptscriptfont1
          \textfont2=\scriptfont2
          \scriptfont2=\scriptscriptfont2
          \textfont3=\scriptfont3
          \scriptfont3=\scriptscriptfont3
        \renewcommand{\arraystretch}{0.7}
        \begin{array}{c}}{\end{array}}
 
\def\wspec{w'_{\rm special}}
\def\mup{\widehat\mu^+}
\def\mupm{\widehat\mu^{+|-_\Lambda}}
\def\pip{\widehat\pi^+}
\def\pipm{\widehat\pi^{+|-_\Lambda}}
\def\ind{{\rm I}}
\def\const{{\rm const}}
\def\ft#1{\widehat{#1}} 
\def\vv#1{\vec{#1}} 
\def\tends#1{\mathop{\longrightarrow}\limits_{#1}}
\def\ssum{\mathop{\hbox{``}\sum\nolimits\hbox{''}}\limits}
\def\hyp#1#2{\smallskip\par\noindent{\bf (#1)  #2}.}

\bibliographystyle{unsrt}
 
%\title{\vspace*{-2.4cm}
 
\title{\vspace*{-2.4cm} Comment on: Critical behavior of the \break
spin diluted 2D Ising model: A grand ensemble approach\break
by R. K\"uhn}

\author{
  \\
  {\normalsize Aernout C. D. van Enter}        \\[-1.5mm]
  {\normalsize\it Institute for Theoretical Physics, Rijksuniversiteit Groningen}   \\[-1.5mm]
 % {\normalsize\it Rijksuniversiteit Groningen}         \\[-1.5mm]
  {\normalsize\it Nijenborgh 4, Groningen, the Netherlands}                \\[-1.5mm]
%  {\normalsize\it 9747 AG Groningen, the Netherlands}           \\[-1.5mm]
%  {\normalsize\it THE NETHERLANDS}             \\[-1mm]
  {\normalsize\tt aenter@phys.rug.nl}        \\[-1mm]
%  {\protect\makebox[5in]{\quad}}  % To force authors' names to be written
                                  %   vertically, one above another.
                                  % (\author seems to put them side-by-side
                                  %   if there is room.)
  {\normalsize C. K\"ulske} \\[-1.5mm] 
%    }\\%[1mm]
{\normalsize\it WIAS, Mohrenstrasse 39, Berlin, Germany }\\[-1.5mm]
%{\normalsize\it Mohrenstrasse 39, }\\[-1.5mm]
%{\normalsize\it Av.\ Prof.\ Luciano Gualberto, }\\[-1.5mm]
%{\normalsize\it Travessa J, 374 T\'erreo} \\[-1.5mm]
%{\normalsize\it Berlin, Germany}\\[-1mm]
{\normalsize\tt kuelske@wias-berlin.de}\\[-2mm]
%}
%  {\protect\makebox[5in]{\quad}}  % To force authors' names to be written
                                  %   vertically, one above another.
                                  % (\author seems to put them side-by-side
                                  %   if there is room.)
  \\[-2mm]
{\normalsize C. Maes} \\[-1.5mm]
{\normalsize\it Instituut voor Theoretische Fysica, K.U.Leuven., }\\[-1.5mm]
{\normalsize\it Celestijnenlaan 200D, Leuven, Belgium}\\[-1.5mm]
%{\normalsize\it Leuven, Belgium }\\[-1mm]
{\normalsize\tt Christian.Maes@fys.kuleuven.be}\\[-2mm]
}

\date{\datetitle}
 
\maketitle
\thispagestyle{empty}

\clearpage

\renewcommand{\baselinestretch}{2}
\small\normalsize

\setcounter{page}{1}
 
%\begin{abstract}

%\end{abstract}
\medskip


%%%%%%%%%%%%%%%%LATEX FORMAT %%%%%%%%%%%%%%%%%%%%%%%%%%%%%%
%\documentclass[10pt]{article}

%\textheight 21.6 truecm
%\textwidth 15.2 truecm
%\oddsidemargin 10truemm

% Macros added
%\def\block{{\rm block}}
%\def\blocks{{\rm blocks}}

%\begin{document}

%\noindent{\Large \bf
%Title
%Comment on: Critical behavior of the randomly %\break
%spin diluted 2D Ising model:
%\break
%A grand ensemble approach, by R. K{\"u}hn,
%\break
%Physical Review Letters 73, 2269 (1994).
\normalsize 
%{\rm}

\vspace{24pt}

%\rm

%\vspace{24pt}
%\footnotesize
%\vspace{3pt}

\normalsize

\vspace{12pt}

%As a direct consequence of some of our recent work, \cite{EMSS} and
%\cite{kue}, it follows 
In this Comment we want to point out 
that the grand ensemble approach applied in
\cite{kuhn} suffers from being ill-defined on the model under
consideration. 
%This was shown in \cite{EMSS} for the diluted Ising
%model studied in \cite{kuhn}, and afterwards in much greater
%generality in \cite{kue}.

We remind the reader that the grand ensemble approach which apparently
goes back to Morita \cite{Mor} consists in rewriting the weights in the
quenched average
\begin{equation}
P(\underbar n, \sigma)= P(\underbar n) {1 \over Z_{\underbar n}}
\exp-H_{\underbar n}(\sigma)
\end{equation}
as Gibbsian weights $\exp -  H^{\phi}(\underbar n,
\sigma)/Z^{\phi}$ in an annealed average for some effective
Hamiltonian $H^{\phi}(\underbar n,\sigma)$. Here the $\underbar n$
denote the occupation number variables, which are 0 or 1
independently on each site with a certain prescribed dilution
probability, and the $\sigma$ denote Ising spins, which are present
on occupied sites, and interact at inverse temperature $\beta$ via 
a nearest-neighbor interaction.

The "disorder potential" $\phi$ describes the difference
between the original Hamiltonian and this effective Hamiltonian.

\smallskip

%The main result of the works mentioned above is that 
The assumption that  an effective Hamiltonian exists for some
given distribution (measure ) is not an innocent one
as has been known for some time \cite{EFS1} .
Indeed, recently (\cite{EMSS} and \cite{kue}) it was proven 
that in the thermodynamic
limit there does {\it not} exist a well-behaved interaction potential, 
describing
such an effective disorder-potential Hamiltonian. In other
words, due to severe nonlocalities, these quenched measures are non-Gibbsian,
for the model of K\"uhn  \cite{EMSS}, as well as for more general disordered
models \cite{kue}, just as in \cite{EFS1} various renormalized measures 
were shown to be. 
%From \cite{EFS1}it became clear that the assumption that 
%an effective Hamiltonian for some measure exists is in general 
%not an innocent one.

We emphasize that this result occurs at low temperatures,
but at arbitrary dilution. Hence critical points, as
well as open regions in the dilution density-temperature plane around them, of
the model studied in \cite{kuhn} are certainly affected.
Thus the approximations used in \cite{kuhn} are intrinsically uncontrolled.

How reliable the conclusions reached in \cite{kuhn} are, remains therefore
to be seen.

On the negative side, nonlocalities can of course strongly influence
long-range properties, and critical properties are preeminently long-range
properties.

On a more positive side, as in various other examples 
\cite{EMS},
one can show very generally 
\cite{kue2} that these quenched measures belong to the "weakly
Gibbsian" class, cf. \cite{BKL, dobsh2, EMS, MRSV}. 
%Informally this
%means that, 
%Although there does not exist a "nice" interaction
%potential, a "not so nice" interaction potential can be found. In
%this case the system's distribution can be described via an
%effective Gibbsian interaction, with relative energies well-defined
%on a full measure set of configurations, but not uniformly
%summable. 
%changes 
 Moreover, for the ferromagnetic Gibbs 
state, there is really an expansion of the (almost surely defined)
interaction potential 
%(with this property) 
in terms  of the form $\lambda_P \prod_{i\in P}n_i$
where $P$ is running over the 
connected plaquettes on the lattice (as 
was used in \cite{kuhn}).  
%This form is not to be taken for granted for any Gibbs state 
%of the system: There is no 
Such an expansion does not always exist; for the random 
Dobrushin-state for example it does not (although an expansion
of a different form does exist)\cite{kue2}. 

% end of changes

It might be that this Gibbsian restoration of
non-Gibbsian states (as carried out explicitly in e.g. \cite{MRSV})
can to some extent explain that, as with renormalization group computations, 
often the results obtained by a priori
mathematically objectionable methods turn out to be surprisingly
good.

We claim that our results go some way in meeting the desire
expressed  in \cite{kuhn} that " a deeper understanding of our
approach would... be welcome".


\begin{thebibliography}{99}

\bibitem{BKL}
J.~Bricmont, A.~Kupiainen and R.~Lefevere,
\newblock
Renormalization group pathologies and the definition of Gibbs
states.
\newblock
 {\em  Comm. Math. Phys.}, 194:359--388, 1998.


\bibitem{dobsh2}
R.~L.~Dobrushin and S.~B.~Shlosman
\newblock ``Non-Gibbsian'' states and their Gibbs description.
\newblock {\em Comm. Math. Phys.}, 200:125--179, 1999.

\bibitem{kuhn}
R.~K{\"u}hn.
\newblock Critical behavior of the randomly spin diluted 2D Ising model: A
grand ensemble approach.
\newblock {\em Phys. Rev. Lett.}, 73:2268--2271, 1994.

\bibitem{kue}
C.~K{\"u}lske.
\newblock Non-Gibbsianness and phase transitions in random lattice spin models.
\newblock {\em Markov Proc. and Related Fields}, 5: 357-388, 1999

%changes 
\bibitem{kue2}
C.~K{\"u}lske.
\newblock Weakly Gibbsian representations for joint measures of 
quenched lattice spin models, available at  
{ http://www.ma.utexas.edu/mp{\--}arc/ } preprint 99-411.
%wias-berlin.de/publications/preprints, 1999
%end of changes

\bibitem{MRSV}
C.~Maes, F.~Redig, S.~Shlosman and A. Van Moffaert
\newblock
Percolation, Path Large Deviations
    and Weak Gibbsianity.
    \newblock {\em Comm. Math. Phys.}, 208: 517--545, 2000.


\bibitem{Mor}
T.~Morita.
\newblock Statistical Mechanics of quenched solid solutions with application
to magnetically dilute alloys.
\newblock {\em J. Math. Phys}, 5:1402--1405, 1964.

\bibitem{EMSS}
A.~C.~D.~van Enter, C.~Maes, R.~H.~Schonmann and S.~B.~Shlosman.
\newblock The Griffiths singularity random field.
\newblock Dobrushin memorial volume, Am. Math. Soc., to appear.






\bibitem{EFS1}
A.~C.~D. van Enter, R.~Fern{\'a}ndez, and A.~D. Sokal.
\newblock Renormalization transformations in the vicinity of first-order phase
  transitions: {W}hat can and cannot go wrong.
\newblock {\em Phys. Rev. Lett.}, 66:3253--3256, 1991,
 and
\newblock Regularity properties and pathologies of position-space
  renormalization-group transformations: Scope and limitations of {G}ibbsian
  theory.
\newblock {\em J. Stat. Phys.}, 72:879--1167, 1993, and  \break
A.~C.~D.~van Enter and R.~Fern{\'a}ndez. 
\newblock Problems with the definition of momentum-space renormalization 
transformations. 
\newblock {\em Phys. Rev.} E, 59:5165--5171, 1999.

\bibitem{EMS}
A.~C.~D. van Enter, C. Maes and S.~B. Shlosman.
\newblock Dobrushin's program on Gibbsianity restoration: Weakly Gibbsian
and almost Gibbsian fields.
\newblock Dobrushin memorial volume, Am. Math. Soc., to appear.

%\bibitem{ES}
%A.~C.~D.~van Enter and S.~B.~Shlosman.
%\newblock (Almost) Gibbsian description of the sign field of SOS fields.
%\newblock {\em J. Stat. Phys.}, 92:353--368, 1998.



\end{thebibliography}
\end{document}